\documentclass[11pt]{article}  
\usepackage{graphicx}  %
\usepackage[numbers,sort]{natbib}
\usepackage{float}
\usepackage{amsmath}  
\usepackage{amssymb}  
\usepackage{hyperref}
\usepackage{authblk}
\usepackage[ruled,vlined]{algorithm2e}

\title{Engineering Emergence}

\author[1,2,3]{Abel Jansma\thanks{\texttt{a.a.a.jansma@uva.nl}}}
\affil[1]{Dutch Institute for Emergent Phenomena, the Netherlands}
\affil[2]{Institute for Logic, Language and Computation, and}
\affil[3]{Institute of Physics, University of Amsterdam, the Netherlands}

\author[4]{Erik Hoel\thanks{\texttt{erik.hoel@tufts.edu}}}
\affil[4]{Allen Discovery Center, Tufts University, Medford, MA, USA}

\begin{document}

\maketitle

\begin{abstract}

A defining property of complex systems is that they have multiscale structure. How does this multiscale structure come about? We argue that within systems there emerges a hierarchy of scales that contribute to a system's causal workings. An intuitive example is how a computer can be described at the level of its hardware circuitry (its microscale) but also its machine code (a mesoscale) and all the way up at its operating system (its macroscale). Here we show that even simple systems possess this kind of emergent hierarchy, which usually forms over only a small subset of the super-exponentially many possible scales of description. To capture this formally, we extend the theory of causal emergence (version 2.0) so as to analyze how causal contributions span the full multiscale structure of a system. Our analysis reveals that systems can be classified along a taxonomy of emergence, such as being either “top-heavy" or “bottom-heavy" in their causal workings. From this new taxonomy of emergence, we derive a measure of complexity based on a literal notion of “scale-freeness" (here, when causation is spread equally across the scales of a system) and compare this to the standard network science definition of  “scale-freeness" based on degree distribution, showing the two are closely related. Finally, we demonstrate the ability to engineer not just the degree of emergence in a system, but to control it with pinpoint precision.

\end{abstract}

\section{The paradox of scale}

A central reason why complex systems are considered complex is because they have multiscale structure \cite{bar2019dynamics, simon2012architecture}. However, this multiscale structure is regularly hidden, simply because when scientists attempt to understand how a system works, they will often write down the equivalent of a single diagram—such as a causal model (e.g., a structural equation model or directed acyclic graph \cite{pearl2009causality}), or a discrete state machine (e.g., a Markov chain). Writing down only a single set of variables, and charting their influence on each other, is certainly natural \cite{pearl2018book}. Yet this necessarily limits the analysis to a single scale of description (consisting of just those variables, or elements, or mechanisms, or states—depending on the model), which in turn hides a great deal of structure: All the other scales that are being ignored in favor of the one. Descriptions of systems at a single scale are like taking a 2D slice of a higher dimensional object and presenting that slice as if it were the object itself.

\begin{figure}[H]
    \centering
    \includegraphics[width=1\textwidth, height=0.8\textheight, keepaspectratio]{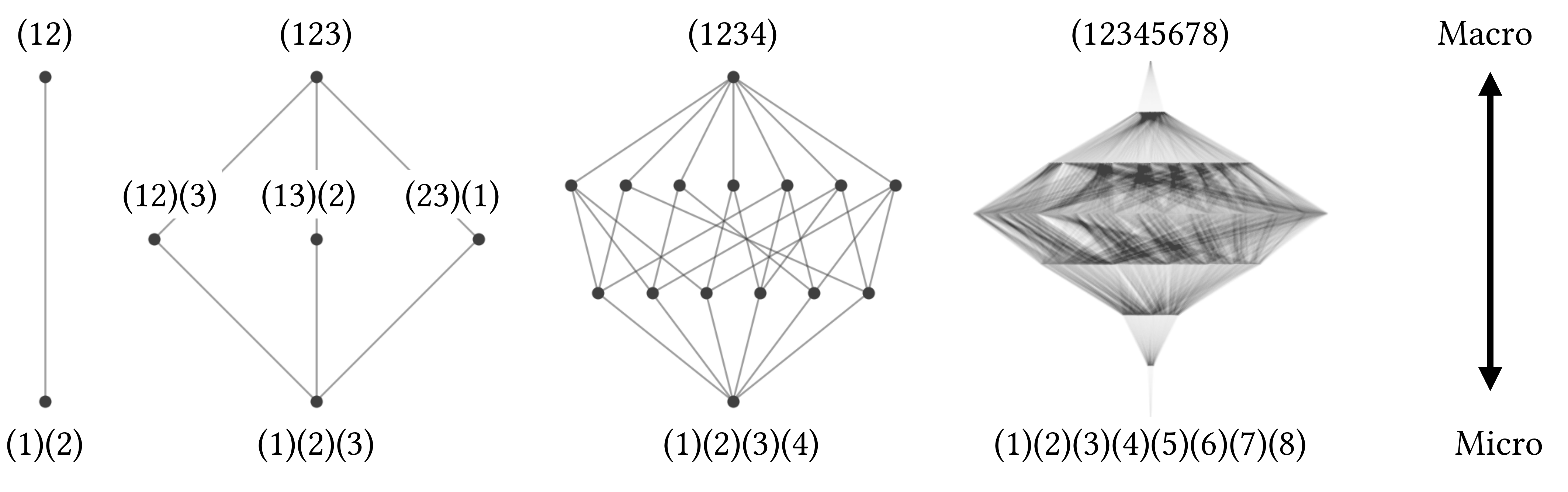}
    \caption{The multiscale structure of a system is based on a partitioning of its states. Partitions can be partially ordered by \textit{refinement} to construct partitions lattices. The associated Hasse diagrams are shown here for 2, 3, 4, and 8-state systems. For any two partitions $\pi_a, \pi_b$ we set $\pi_a \leq \pi_b \iff \forall b \in \pi_b :\exists a \in \pi_a \text{ s.t. } a\subseteq b $. This puts coarse partitions with few blocks (macroscales) at the top, and fine partitions with many blocks (microscales) at the bottom. The number of possible partitions of $n$ states (the number of elements in the unpartitioned microscale at the bottom) grows superexponentially with $n$, namely, as the $n$th Bell number (the lattices above contain respectively 2, 5, 15, 4140 partitions). \label{fig:partition_lattices}}
\end{figure}

One reason this is common is that, when confronted with the full set of possible scales of description, it is a set so astronomically large it at first seems of little use, if not outright confusing. Even for a system with just 8 discrete states, the set of all possible partitions of those states is 4,140 (see Figure \ref{fig:partition_lattices}). This set forms a partially ordered lattice, or \textit{mereology} \cite{jansma2025mereological}, wherein some partitions are finer, or coarser, than other partitions. The ordered lattice can also be thought of as many paths that increase coarseness, spanning from the microscale of the system (its most fine-grained possible description) to the ultimate macroscale (its most coarse-grained possible description). Each horizontal slice of the lattice corresponds to some particular dimensionality (or level) of the system, usually with many possible partitions at each level. 

Notably, the spatiotemporal hierarchy of the sciences (from microphysics to macroeconomics) suggests that among the astronomical number of ways to partition a system, there is a much smaller subset of actually causally-relevant scales that best (or better) explain a system's workings \cite{hoel2024world}, at least when compared to random scales. This is true for individual systems too: e.g., the causally-relevant scales of a computer can also be described sensibly as a hierarchy, with its physical logic gates as a microscale, its machine code as a mesoscale, and its operating system as a macroscale. 

A system can differ in important ways depending on what scale it is being described at. E.g., the mechanisms of a system, such as the kind of logic gates the system contains, can change at different scales of descriptions \cite{hoel2013quantifying}, and consequentially so can the types of computations \cite{gomez2022slicing, bongard2023there}. Broadly, how a system functions can be described at different scales; e.g., brain function may be modeled as a large set of interacting individual neurons, but also as a (smaller) set of interacting cortical minicolumns or brain regions \cite{yuste2015neuron}. 

We are certainly not the first to notice that complex systems have multiple descriptions at different scales: the observation has a long tradition in disparate fields, particularly in cognitive sciences (like Marr's three levels of analysis \cite{marr2010vision}); it also extends to questions of how software arises from hardware \cite{rosas2024software}, or how macroscales are dynamically related to their microscales \cite{barnett2023dynamical}. Additionally, there already exist various information-theoretic proposals for methods to identify particular macroscales that may operate as especially informative compressions of microscale dynamics \cite{gershenson2012complexity, marchese2022detecting}. However, if macroscales add nothing beyond successful compression, this suggests they are not truly causally-relevant at all, and useful only for an observer's convenience.

When analyzing the causal workings of a system directly, the astronomical size of the set of scales creates a seeming paradox when it comes to understanding “What does what?", i.e., how the system actually works. The existence of the full set of scales entails that everything that happens in the system is massively overdetermined \cite{kim2000mind, bontly2002supervenience} (to put it in lay terms: for any one event, there is an astronomically large set of possible explanations across the full set of possible scales). 

Confronted with this embarrassment of multiplicity, one option is to assume relativism, i.e., that when it comes to complex systems, there is “no privileged level of causation" \cite{noble2012theory}. However, this cannot account for the fact that some scales seem to truly capture a system's workings better than others, much like how the many special sciences beyond physics seem to give unique insight into the phenomena they study \cite{fodor1980special}. An alternative but equally extreme option is to assume reductionism, i.e., that only the microscale truly matters for the system's causal workings \cite{block2003causal}. However, this too leads to absurdities (no higher level truly \textit{does} anything)—indeed, this would indicate that most elements of science, the majority of which are macroscales of some sort, cannot truly add anything to a system's workings, and so are epiphenomenal \cite{hoel2018agent}. 

\section{How systems get their scales}

Herein, we demonstrate a framework for resolving this paradox that enables a non-reductive, but also non-mysterious, rich view of multiscale structure. Specifically, we propose a method for identifying the subset of scales (defined as coarse-grainings of a system based on partitions) that contribute—non-redundantly and uniquely—to the system's overall causal workings. We judge this subset of scales to be causally emergent \cite{hoel2025causal}, since the causation they contribute cannot be fully reduced to any underlying scale (i.e., any scale below them along any micro$\rightarrow$macro path).

To establish this framework, first we note that we can formalize each individual partition (such as those in Figure \ref{fig:partition_lattices}) as representing an individual scale, which is expressible as a causal model that describes the system's workings. Here, we focus on systems defined by transition probability matrices (TPMs) that reflect what the system would do (how it would transition next) in any given state $X_t$, or under any given intervention possible in the statespace (such as $do(X_t=x)$). 

Normally, identifying causal models involves inference under a myriad of difficulties and limitations. For simplicity, both conceptual and mathematical, we therefore confine our analysis to causal models that are already known in their entirety, and are represented as Markov chains (as described by a given TPM—see SI \ref{SI:TPM_do_calc} for details). We also assume that a system's microscale is well-defined, in that it forms a specific TPM, and that all the true conditional probabilities (the state transitions) are a priori known. For a system that meets those conditions, this still results in an enormous set of TPMs representing scales beyond the microscale, any one of which are on the way to other (more macroscale) TPMs (that is, they lie along some micro$\rightarrow$macro path). This can be seen by coarse-graining the original microscale TPM of a system according any one of its many possible partitions, such that, e.g., for partition $(123)(456)$ the microstates $1,2,3$ and $4,5,6$ would be coarse-grained together, and the probabilities of their possible effects (next states) and their possible causes (previous states) would also be summarized together, resulting in a macroscale TPM that has only two macrostates (see \cite{hoel2013quantifying, klein2020emergence} for more details and methods on coarse-graining microscale TPMs into macroscale TPMs).

Some of these many TPMs, each corresponding to a scale of description, are informative about the causal workings of the system as it moves from $t$ to $t+1$. Most are not. This is because a randomly chosen scale, and the constructed TPM it specifies, is often junk: by which we mean that the conditional probabilities are garbled and noisy. For a randomly chosen scale, upon examining its TPM that defines its dynamics from $t$ to $t+1$, causes do not deterministically lead to effects (in the form of future states), and for any given effect, there are many possible common causes (previous states). One can think of this problem intuitively as: “What would happen if you randomly coarse-grained parts of a computer together?"—most such coarse-grainings will average over random parts from different transistors together, resulting in obscure variables and confusing relationships. Conversely, for a much smaller subset of scales, their TPMs are intuitively informative and interesting, and bring the higher-scale causal relationships of the system into focus.

Specifically, we can judge the causal relationships at a particular scale of the system (again, defined as a TPM) based on interventions of the type $do(X_t=x)$, performed at some time $t$ , and their effects observed at $t+1$. For causal relationships to be strong (a property which is sometimes called “powerful," or “informative," or other synonyms) they must be deterministic (individual interventions should have few possible effects), and non-degenerate (most effects should be unique). 

Assuming the simple case of Markovian systems, fully characterized by a TPM that captures the conditional probability of the state of the system at $t+1$ (which we refer to as an effect $e$), given the state of the system at time $t$ (which we refer to as a cause $c$): $T_{ce}=p(e|c)$, then \textit{determinism} is defined based on some individual cause $c$ (here, a state in a TPM), and a possible set of future effects (here, over the set of states possible at the next timestep).

\begin{align}
\text{determinism}(c) &= 1 + \frac{1}{\log_2 n} \sum_{e \in E} p(e|c)\log_2 p(e|c)\\
&= 1 - \frac{H(E|c)}{\log_2 n}
\end{align}

Being based on the entropy, $H$, it is an information-theoretic quality that captures the degree of certainty about effects, given a cause, and is therefore sensitive to the strength of the conditional probabilities of a system \cite{hoel2013quantifying}. Herein, we will refer only to the system-wide determinism, which is calculated based on the average occurrence of any given transition from $c$ to $e$ (a system transition from $t$ to $t_{+1}$), given some prior distribution over causes, or interventions, $p(C)$, for which we use a uniform distribution at each scale \cite{comolatti2025consilience}. For a system with transition probability matrix $T$, the cause-specific and system-wide determinism are given by
\begin{align}
    \text{determinism}_T(c) &= 1 + \frac{1}{\log_2 n} \sum_{e \in E} T_{ec}\log_2 T_{ec}\\
    \text{determinism}_T &= \mathbb{E}_{p(C)}\left[ \text{determinism}_T(c) \right]\\
    &= 1 + \frac{1}{\log_2 n} \sum_{c \in C} \sum_{e \in E} p(c) T_{ec}\log_2 T_{ec}\\
    &= 1 - \frac{H(E|C)}{\log_2 n}
\end{align}

Conversely, \textit{degeneracy} captures the degree of overlap in the effects of causes, given a prior $p(C)$, calculated as:
\begin{align}
    \text{degeneracy}_T &= 1 - \frac{H(E)}{\log_2 (n)}\\
     &= 1 + \frac{1}{\log_2 n} \sum_{e \in E} T_e\log_2 T_e
    \intertext{where}
    T_e &= \sum_{c \in C} p(c) T_{ec}
\end{align}
Determinism is $1$ and degeneracy is $0$ only in cases of a TPM that is a permutation matrix; determinism is $1 $ and degeneracy is $1$ when all states of a TPM transition to the same single state (all causes have the same common effect); and both are $0$ when the transitions of a TPM are completely random. For this reason, determinism and degeneracy are closely related to the dynamical reversibility \cite{zhang2025dynamical}. Additionally, the two terms are also closely related to the axiomatic definition of causation \cite{hoel2025causal}, being an information-theoretic generalization of the concepts of the sufficiency and necessity, respectively; when causes are always completely sufficient for their effects, determinism is high, and when causes are always necessary for their effects, degeneracy is low. 

Measures of probabilistic causation, which capture to what degree something is a cause, \cite{fitelson2011probabilistic}, are generally reducible to the sufficiency and necessity, and therefore are closely related to the determinism and degeneracy as well \cite{comolatti2025consilience}. Therefore, we refer to the determinism and the degeneracy together as the \textit{causal primitives} ($\text{CP}$) of a scale, as in CE 2.0 \cite{hoel2025causal}, to reflect that a well-tuned chosen measure of causation (like a measure of causal power, or strength, or degree, or informativeness, or constraint, or some other interpretation) will set some analog of these two terms in relationship, such that causes are more powerful (or stronger, or more informative, etc.) when determinism is closer to $1$ and degeneracy is closer to $0$. 

For simplicity's sake, we define the inverse of the degeneracy ($1-degeneracy$) as the \textit{specificity}. Then, we capture the two terms' relationship by simply adding the two terms (determinism and the inverse of degeneracy) together, referring to this summation as $\text{CP}$, and this is calculated at each scale of the system (i.e., for each TPM), in a way that normalizes to a $[0,1]$ range:
\begin{align}
    CP = \text{determinism}_T + \text{specificity}_T-1 \label{eq:EI}.
\end{align} 
$\text{CP}$ represents a particular scale's contribution to the system's causal workings, and is equivalent to the \textit{effectiveness} defined in \cite{hoel2013quantifying}.

\begin{figure}[H]
    \centering
    \includegraphics[width=1\textwidth, height=0.8\textheight, keepaspectratio]{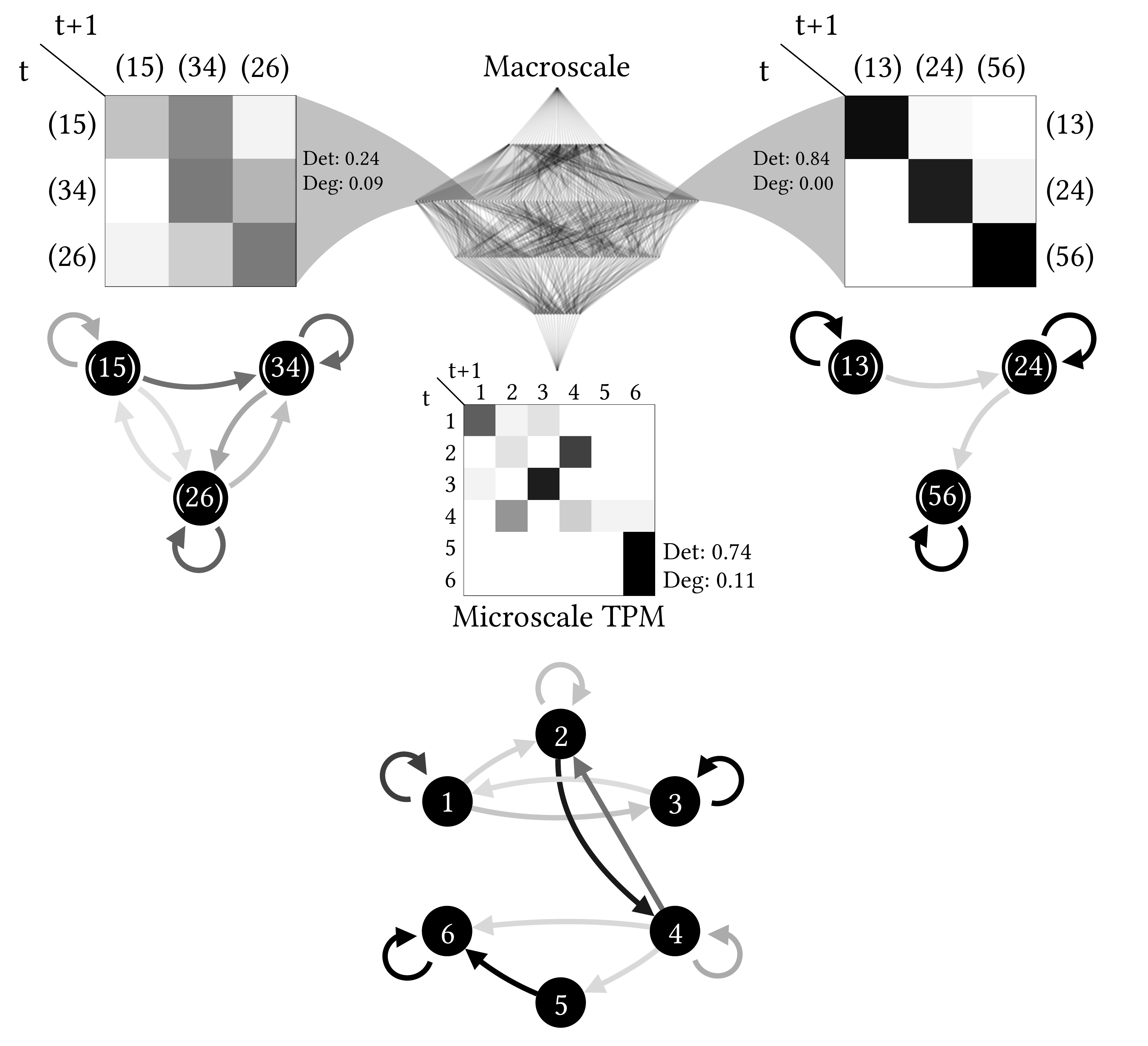}
    \caption{A demonstration of calculating $\text{CP}$ at specific scales out of the full set of scales of a 6-state system (here a Markov chain described by its microscale TPM, the probabilities of which are shown in grayscale in the middle). Above the microscale TPM we show the associated Hasse diagram. Moving up the lattice traverses all coarse-grainings, proceeding from smaller groups to larger one. \emph{Left:} a generic coarse-graining that reduces  $\text{CP}$ (by lowering determinism) indicating that scale fails to add anything to the system's causal workings. \emph{Right:} an especially informative coarse-graining that groups the six states into appropriate pairs, increasing determinism and decreasing degeneracy. This is reflected in the associated graph by the dominance of self-loops.\label{fig:coarse_grainings}}
\end{figure}

An example of calculating the combined causal primitives for different scales of a system is shown in Figure \ref{fig:coarse_grainings}. Two scales out of the set of all possible scales of a system were chosen at the same level (the same degree of dimension reduction), and each were used to construct two different TPMs. Notably, the two respectively differ significantly in their determinism and degeneracy.

The microscale TPM has a $\text{CP}$ value of 0.63. Meanwhile, the macroscale TPM highlighted on the left in Figure \ref{fig:coarse_grainings} has a $\text{CP}$ value of only 0.13, reflecting that the causal structure is represented in a noisy manner, with transitions that are highly uncertain. Comparatively, the macroscale TPM on the right has a $\text{CP}$ value of 0.84, reflecting a positive gain in $\text{CP}$, because the causal relationships are more deterministic and less degenerate at that scale, compared to the underlying microscale. That is, while the scales themselves are reducible to their underlying scales (being a coarse-graining of the microscale), they can still add to the system's causal workings (here reflected by the joint $\text{CP}$ value) by increasing the determinism and decreasing the degeneracy. Mathematically, this is like adding error correction to causal relationships due to the multiple-realizability of macroscales \cite{hoel2018agent}.

Note that measuring these quantities (the system-wide determinism and degeneracy) requires a prior distribution over causes $p(C)$ to reconstruct the joint distribution from $T$. This corresponds to an \emph{intervention distribution} that specifies the probabilities with which each of the interventions are performed at this scale (it can also be thought of as the set of possible counterfactuals in the causal analysis). It is important that $p(C)$ is not solely based on the observed distribution, since in most cases interventions/counterfactuals are required for capturing causation \cite{pearl2009causality} (here we use a uniform distribution, to reflect treating counterfactuals/interventions equally—see the Discussion for more details).

\section{How hierarchies emerge across scales}

Previous research introducing CE 2.0 examined how $\text{CP}$ could increase across a single micro$\rightarrow$macro path \cite{hoel2025causal}. Here, we extend CE 2.0 to be calculated over all possible micro$\rightarrow$macro paths that traverse a system: that is, we demonstrate how to distribute the determinism and degeneracy of a system out across the full set of scales. This carves away all but the subset of scales that non-redundantly contribute to the system's causal workings: this remainder we call the \textit{emergent causal hierarchy} (or\textit{ emergent hierarchy} for short), defined as the subset of scales that contribute non-zero and non-redundant gains in $\text{CP}$. We introduce a taxonomy for these emergent hierarchies: some are “top-heavy," others “bottom-heavy," and others form peaks at mesoscales. 

This taxonomy can be quantified by further information-theoretic measures that capture how complex the emergence in a system is, based on how causal contributions are spread across scales. Essentially, we introduce a literal notion of “scale-freeness" wherein if no one scale dominates the causal workings of a system, it has a high degree of emergent complexity. We also demonstrate how shifting network growth rules can in turn shift a system's emergent hierarchy from “bottom" to “top-heavy," and this reveals how our notion of emergent complexity overlaps with the traditional notion of “scale-freeness" from network theory. To simulate these networks, and to expand our methods to larger systems, we introduce a heuristic that operates as a branching greedy sampling procedure that sets the stage for performing our analyses on empirical data.

Finally, we explore how our methods can be used to engineer the degree of emergence in a system with pinpoint precision. Specifically, we demonstrate the ability to design systems that possess only a single emergent scale of description in their emergent hierarchy, ensuring that the system's workings are driven by just one lone macroscale and no others.

\subsection{Detecting and visualizing the emergent hierarchy of a system}

\begin{figure}
    \centering
    \includegraphics[width=1\linewidth]{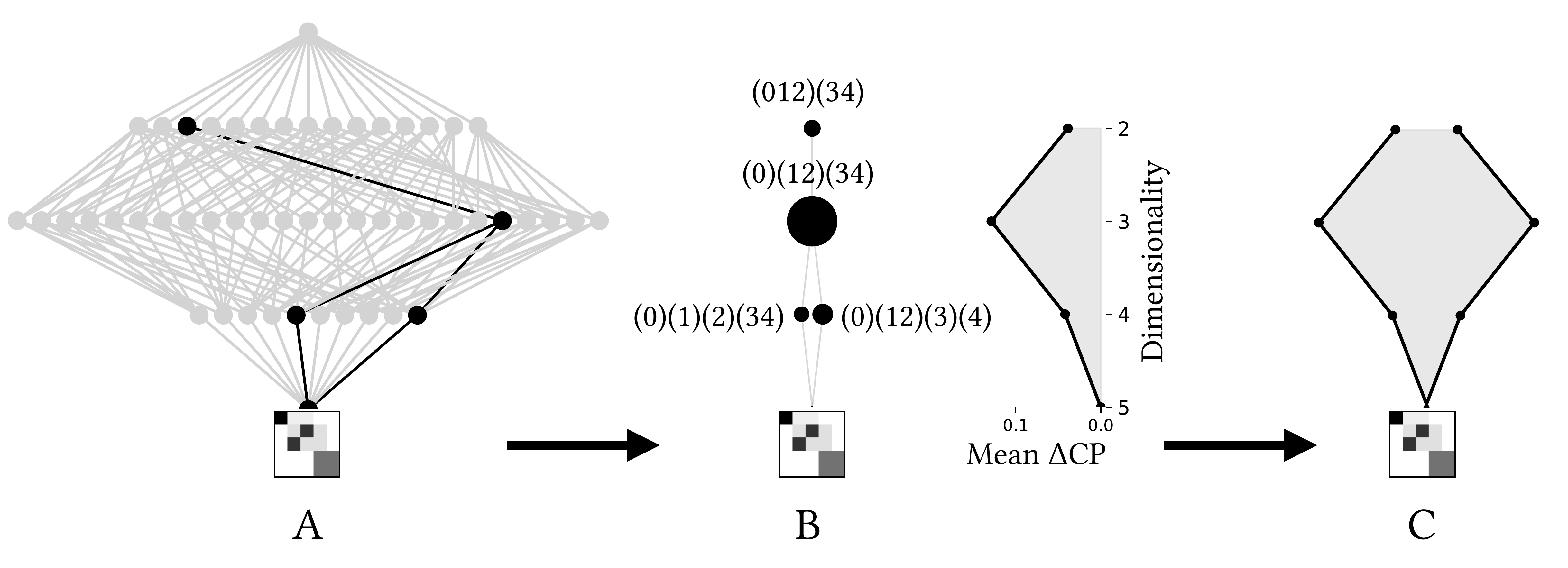}
    \caption{A visualization of our method on a small system with 5 states that correspond, respectively, to a source state, two cycle states, and two sink states. The microscale TPM is shown at the bottom of the panels. Panel A shows the full partition lattice over 5 states, which consists of 52 possible coarse-grainings. Highlighted in black are the scales with a positive $\Delta\text{CP}$ score that increase the CP value beyond any of the scales below them. Panel B shows the sublattice that contains only these five scales (the emergent hierarchy) with the corresponding partitions indicated next to the nodes. In the visualization of the emergent hierarchy, the size of the node reflects its $\Delta\text{CP}$ value. Note that the intuitively “correct" scale that partitioned the system into source, cycle, and sink nodes indeed contributes the most $\Delta\text{CP}$. This is further shown by the vertical plot next to it that shows the average $\Delta\text{CP}$ value per row (per dimensionality of the coarse grained systems). Finally, panel C shows an abstracted representation of the hierarchy, which is just a symmetrized version of the vertical $\Delta\text{CP}$ plot. This representation is used in later figures to illustrate the difference between top-heavy, bottom-heavy, and complex systems. 
    }
    \label{fig:example_system}
\end{figure}

\begin{figure}[H]
    \centering
    \includegraphics[width=1\textwidth, height=0.8\textheight, keepaspectratio]{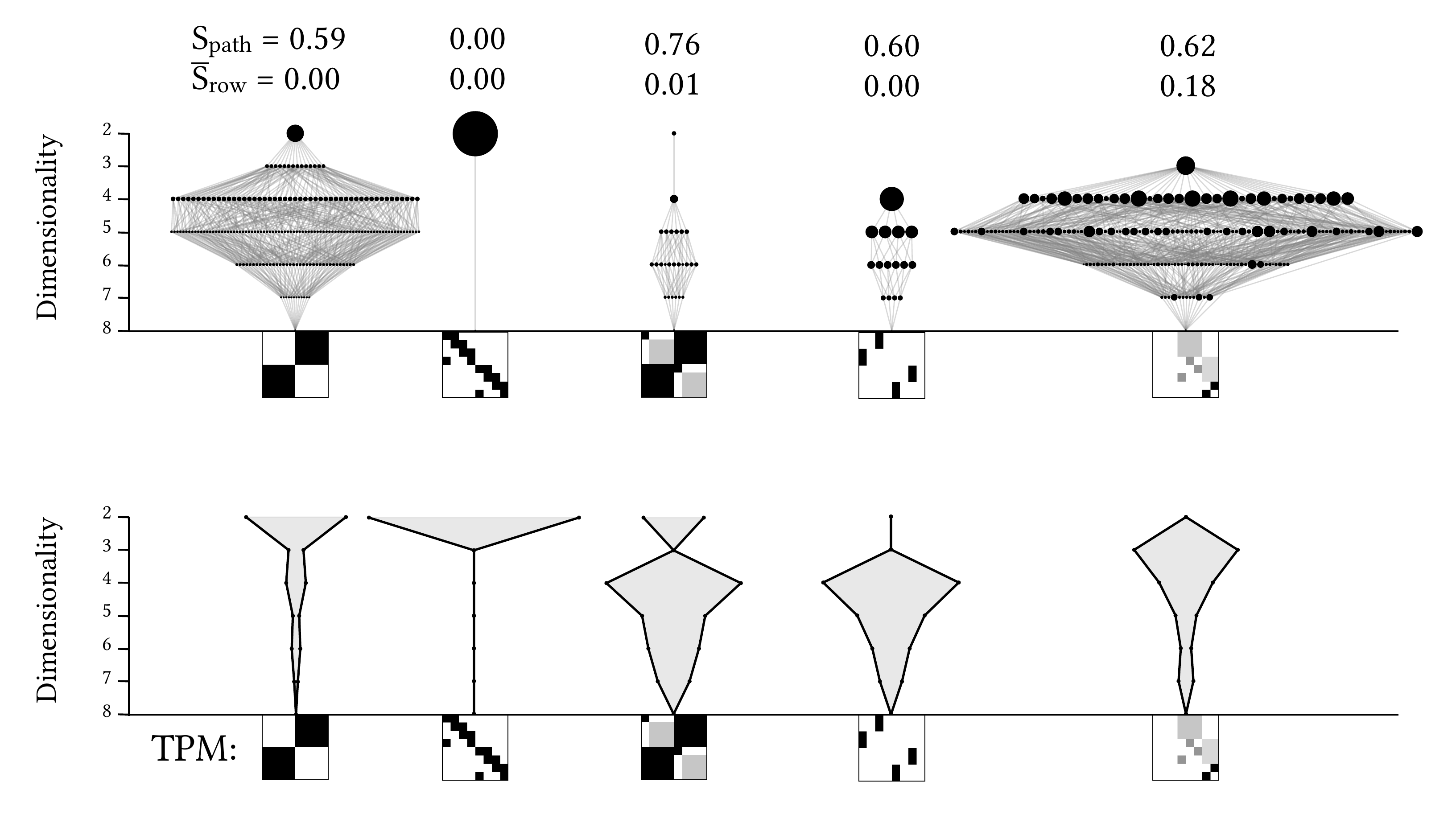}
    \caption{Systems can display causal emergence in many ways, here this is showcased as a “garden" of five systems (all the same size at the microscale) that have different emergent hierarchies. \emph{Top:} The emergent hierarchies visualized across the eight levels of coarse-graining. We show the sublattice of the full partition lattice with positive \(\Delta \mathrm{CP}\), i.e., the nodes that correspond to coarse-grainings that contribute $\text{CP}$ beyond any of its ancestors in the lattice. The size of the node is proportional to \(\Delta \mathrm{CP}\). From left to right, the systems correspond to i) a system composed of two equivalence classes of states, ii) a system with two length-4 cycles (explored in more detail in Figure \ref{fig:balloons}), iii) a system with a mesoscale (taken from \cite{hoel2025causal}), iv) a system with fully deterministic but degenerate transitions, and v) a system with three modules: three source states, a 3-cycle processing module, and 2 alternating sink states. \emph{Bottom:} The width of the shaded region above each TPM corresponds to the average \(\Delta \mathrm{CP}\) at that level (a particular row in the sublattice) of the emergent hierarchy shown in the panel above it (as in Figure \ref{fig:example_system}C) .\label{fig:lattice_garden}}
\end{figure}

Identifying the emergent hierarchy of a system is much like carving away until the underlying causal skeleton that spans scales is revealed. To ensure the clearest conceptual explanation, we first show an overview of our method on a small $5$-state system (Figure \ref{fig:example_system}), and then present some example systems which at the microscale are small enough ($8$ states) that we can calculate via brute force all possible partitions, TPMs, and $\text{CP}$ values, but that nonetheless contain distinct emergent hierarchies (Figure \ref{fig:lattice_garden}).
 
For a system defined by some microscale TPM (the transitions of which are known as ground truth) and with all partitions and their associated TPMs and $\text{CP}$ in hand, we use a \textit{causal apportioning schema} that appropriately distributes out the $\text{CP}$ across all the scales. 

For any given scale, its $\text{CP}$ is compared to all the $\text{CP}$ values of all the scales of the system that are on a path below it in the lattice of scales. Its non-redundant and unique \(\Delta \mathrm{CP}\) value is its positive gain compared to the maximum $\text{CP}$ value of any scale below it, as long as it the comparative scale is along any given micro$\rightarrow$macro path that leads through it (panel A in Figure \ref{fig:example_system}). This is somewhat like a game-theoretic quantity, asking how much each “player" (here, a scale) contributes to the overall game that is the system's workings (see the pseudocode in SI Section 2 for a formal description of this process). The result is the sublattice of remaining scales that have non-zero \(\Delta \mathrm{CP}\) values, which we refer to as the emergent hierarchy (panel B in Figure \ref{fig:example_system}). It is still organized such that partitions of a certain size (a certain dimensionality) are represented in rows on the sublattice (such that for any given dimension, there can be a number of scales that share that same dimensionality, or level). 

While we refer to this procedure as an apportioning schema, it is generally not the case that the sum of all $\Delta$CP values adds up to the maximum attainable $\text{CP}$ of 1; however, we do note that in practice the sum of \(\Delta \mathrm{CP}\) reflects the total possible $\text{CP}$, and that this method has both the advantage of conceptual simplicity and resulting in a clearly distinguishable taxonomy of emergent hierarchies.

The emergent hierarchies for a variety of $8$-state TPMs are shown in Figure \ref{fig:lattice_garden} (top). There is a significant diversity between them: they form a kind of garden of different degrees and types of emergence, the taxonomy of which ranges from “top-heavy" (wherein most $\text{CP}$ is concentrated near the ultimate macroscale), to more “bottom-heavy," to more “mesoscale" (wherein $\text{CP}$ peaks at midway, or thereabouts, along the dimensions of the system); some even look like “balloons" in that there is only a single emergent scale that contributes to the workings of the system. This taxonomy can also be visualized by looking at how the average non-zero \(\Delta \mathrm{CP}\) is distributed across the dimensions or levels of the system, plots of which are shown in Figure \ref{fig:lattice_garden} (bottom). 

\subsection{Complexity as literal scale-freeness}
While the difference between emergent hierarchies can be seen qualitatively, here we introduce two mathematical tools which, in combination, quantify the complexity of an emergent hierarchy.

For the first tool, we follow the intuition that if a system's emergent hierarchy is heavily concentrated at just a few dimensionalities, it must be somewhat simple. This can be assessed by considering the paths that traverse the emergent hierarchy from the bottom microscale to the top macroscale (all the micro$\rightarrow$macro paths). For a system with an $L$-dimensional microscale, these are paths of length $L$, with corresponding $\Delta$CP values $(\Delta\text{CP}_1, \ldots, \Delta\text{CP}_L)$. We normalize these along the path to define:
\begin{align}
p_i \;=\; \frac{\Delta \mathrm{CP}_i}{\sum_{j=1}^{L} \Delta \mathrm{CP}_j}
\quad
\text{for } i = 1,\dots,L
\end{align}
which results in a normalized distribution, $\{p_1,\dots,p_L\}$ over a single path, with associated entropy:
\begin{align}
    H(p) = -\sum_{i=1}^{L} p_i \log_2(p_i).
\end{align}

Note that this is identical to the emergent complexity measure defined for a single micro$\rightarrow$macro path in CE 2.0 \cite{hoel2025causal}. However, as we are extending CE 2.0 to beyond measuring a single path, we take the average of these path entropies across the full set of paths that traverse the emergent hierarchy and define $S_\text{path} = \sum_{p}H(p)$, where the sum is over all paths from the bottom to the top of the emergent hierarchy.

Additionally, now that CE 2.0 can handle multiple paths, there is also more information about how complex the emergent hierarchy is beyond its spread across levels, since there may be multiple scales within a level (e.g., like $(1,2)(3,4)$ and ($1,3)(2,4)$), which share the same dimensionality and occupy the same row on the sublattice of the emergent hierarchy. 

Therefore, we introduce a second way to judge the emergent complexity: how much differentiation there is between the scales at a particular level (i.e., to what degree the \(\Delta \mathrm{CP}\) values vary within rows in the sublattice of the emergent hierarchy).

Let $\Delta^l_i$ denote the $\Delta$CP value for the $i$th $l$-dimensional partition. We normalize this per row as $\bar{\Delta^l_i} = \Delta^l_i / \sum_{j}\Delta^l_{j}$ to define

\begin{align}
    S_l = -\sum_i \bar{\Delta^l_i} \log_2 \bar{\Delta^l_i}
    \intertext{and define the row entropy as its average}
    S_\text{row} = \frac{1}{L}\sum_l S_l
\end{align}

When $S_\text{row}$ is large, then there is a large degree of symmetry between the different scales at that level. To capture complexity, we therefore take its informational inverse, the \textit{row} \emph{negentropy}:
\begin{align}
    \overline{S}_\text{row} = \log_2 L - S_\text{row}\label{eq:row_negentropy}
\end{align}

Thus if the emergent hierarchy has positive $\Delta \mathrm{CP}$ spread across its different levels (large $S_\text{path}$), \textit{and} there is significant differentiation in terms of the size of $\Delta \mathrm{CP}$ within those levels (large $\overline{S}_\text{row}$), then that system possesses a complex emergent hierarchy, and no one scale dominates. 

In Figure \ref{fig:lattice_garden}, for example, the system on the far left and on the far right are both “top-heavy," and therefore their path entropy is similar (shown at the top of the figure for each system). However, the system on the far right has much more differentiation in terms of $\Delta \mathrm{CP}$ between the scales at any given level, and so has greater emergent complexity. In this, the emergent complexity measures how \textit{literally scale-free} a system is in terms of its causal workings.

\subsection{Growing literally scale-free systems}

To demonstrate how our notion of emergent complexity is related to literal scale-freeness, we make use of network growth rules, using the derived networks to construct different microscale TPMs that have different emergent hierarchies with different properties.

Specifically, across a set of growing model networks we parameterize the preferential attachment, $\alpha$, which changes the way a network grows as new nodes are added. For each different $\alpha $, this creates a different network, which here we treated as a directed network representing a set of state transitions (assuming a uniform split of transition probabilities for any node with multiple outputs); i.e., in this interpretation, each node of the network represents a state, and each connection in the network represents a possible transition. In other words, instead of using the growth rule to just grow networks, we use it to grow TPMs by normalizing the adjacency matrix to derive transition probabilities.

During this process, if $\alpha = 1$ then the network is a classic Barabási-Albert network, also known as a  “scale-free" network \cite{barabasi1999emergence}. Scale-free networks are networks wherein connectivities between nodes follow a scale-free power-law distribution, and thus there are a small number of extremely high-connectivity nodes (intuitively, this traditional notion of “scale-freeness" refers how the degree distribution doesn't change when “zooming out").

Such networks have important properties: they are commonly found in the real data and they have a tight link to criticality \cite{aldana2003natural}. However, if during network growth $\alpha > 1$ then the resultant networks have superlinear preferential attachment, and are no longer scale-free, instead congealing into distinct clusters as $\alpha $ increases (eventually forming a starlike TPM where all states transition to the same state and therefore form one large macrostate). The other regime, where $\alpha < 1$, represents sublinear preferential attachment, which forms sparse and branching TPMs.

Differences in growth rules are best seen in larger networks, so therefore we developed a heuristic to reveal the emergent hierarchies of larger networks (this also sets the stage for our methods to eventually be used to analyze real data). Rather than pre-calculating the full set of possible scales and their $\text{CP}$, we developed a “branching greedy algorithm" which can start at the microscale and, by repeatedly traversing the set of system scales in different ways, average over the paths of the system to reveal how \(\Delta \mathrm{CP}\) is distributed (see SI Section 2 for the algorithm's details).
 \begin{figure}[H]
    \centering
    \includegraphics[width=1\textwidth, height=0.8\textheight, keepaspectratio]{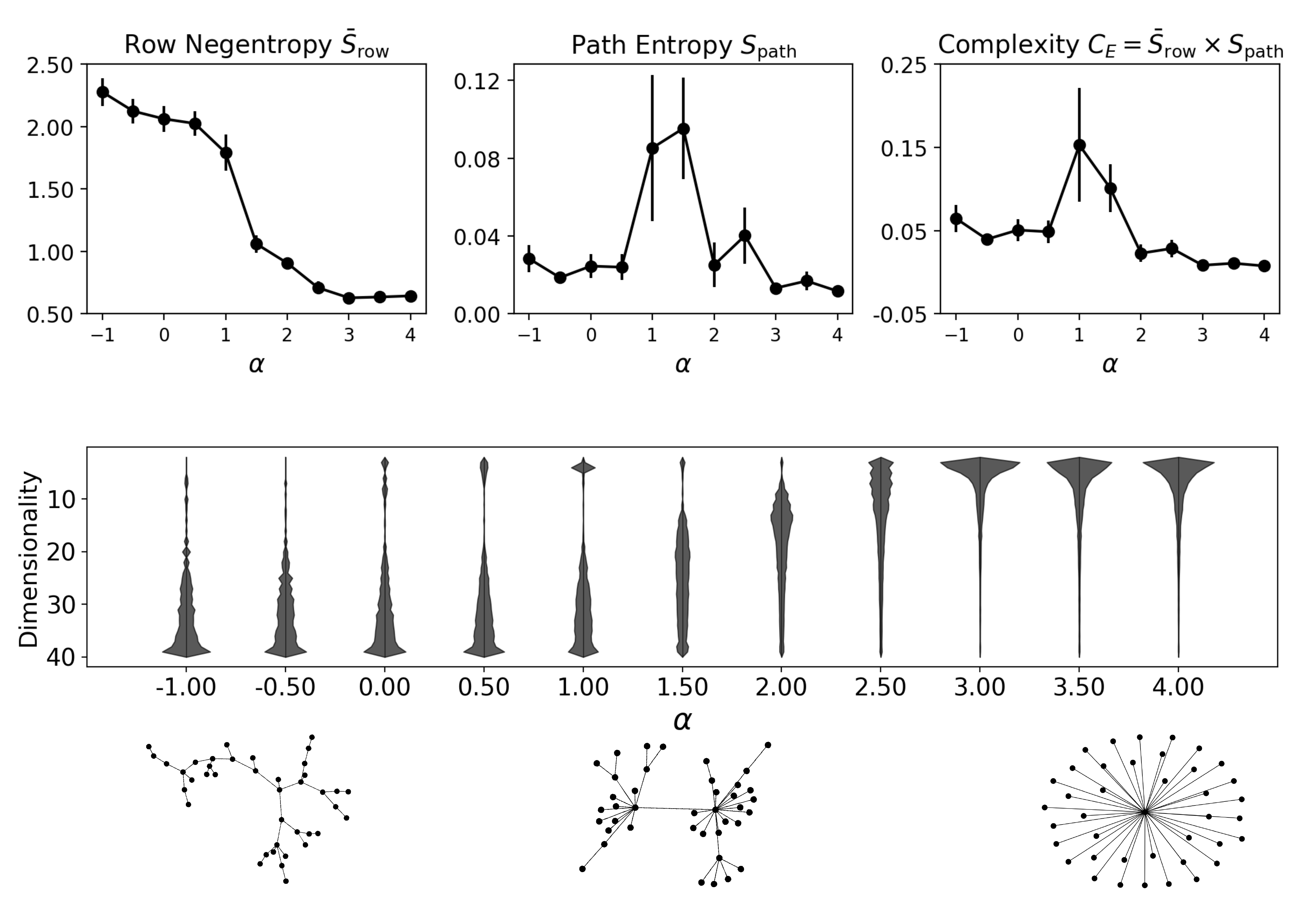}
    \caption{Preferential attachment networks show different emergent hierarchies depending on the preference parameter $\alpha$, and display a phase transition around the $\alpha=1$ scale-free point. The number of connections for newly introduced nodes is set to $m=1$ throughout, in order to yield sparse and more hierarchical tree-like networks. \emph{Top:} We show three measures that contribute to the system's complexity. The average negentropy of \(\Delta \mathrm{CP}\) across the levels of the system (Equation \ref{eq:row_negentropy}) is shown on the left (capturing the differentiation of \(\Delta \mathrm{CP}\) within levels). In the middle are plotted the average entropies of \(\Delta \mathrm{CP}\) across 100 uniformly sampled paths on the sublattice (capturing how uniformly \(\Delta \mathrm{CP}\) is spread across levels). For each $\alpha$ value, we report the mean and standard error across five sampled networks with 40 nodes each. The row negentropy sharply drops beyond $\alpha=1$, which reflects a strong decrease in the differentiation of \(\Delta \mathrm{CP}\) within levels. On the right is plotted the multiplied values of both row negentropy and path entropy. Previous methods, like \cite{hoel2025causal} and \cite{klein2020emergence}, do not take into account parallel paths and therefore couldn't capture this horizontal heterogeneity. \emph{Bottom:} We show the mean $\Delta$CP value for every dimension (in the same representation as in the bottom of Figure \ref{fig:lattice_garden}), averaged across five sampled networks at each $\alpha$. The distribution of $\Delta$CP can be seen to go from “bottom-heavy" at low $\alpha$ to “top-heavy" at high $\alpha$, with the transition taking place between $\alpha=1$ and $\alpha=1.5$. }
\end{figure}

Our results demonstrate how the degree of causal emergence, and its distribution across the emergent hierarchy, changes dramatically under different regimes of preferential attachment. When $\alpha < 1$ the networks are in a bottom-heavy regime, where the scales that causally contribute cluster around the microscale. As $\alpha$ approaches $1$, there is a phase shift wherein causal contribution distributions begin to shift upward (as captured by the increase in the path entropy) and also there is rapid increase in differentiation across the scales at any given level (captured by the negentropy of the distribution of \(\Delta \mathrm{CP}\) at a particular dimensionality, i.e., any row in the sublattice). Together, this means that the emergent complexity peaks at around the regime of $\alpha = 1$. However, in the superlinear regime of attachment, emergent hierarchies become increasingly top-heavy (a low path entropy) and at any given level \(\Delta \mathrm{CP}\) is quite uniform, and so the emergent complexity decreases.

Interestingly, the traditional notion of scale-freeness is linked to criticality \cite{aldana2003natural}, which is in turn tied to properties like information processing and computation capabilities \cite{shew2013functional}, as well as adaptability and evolvability \cite{hidalgo2014information}. However, that traditional definition of “scale-free" is defined based solely on the power-law distribution of connectivity \cite{barabasi1999emergence}. Our results identify an overlapping, but likely not identical, regime wherein networks have emergent hierarchies that are highly complex, spanning the levels of a system and varying within a level. That is, we find that the regime of connectivity wherein emergent hierarchies are complex represents a regime that is \textit{literally} scale-free, in that no one scale of description of the network dominates, instead of just the degree distribution following a power law. 

\subsection{Engineering “pinpoint emergence"}

In the previous section, differently chosen growth rules biased the kind of emergence exemplified by the system, leading to different emergent hierarchies. This indicates that emergence can be steered, and brings up the question: Can one engineer emergence with pinpoint precision?

An intriguing hint is an example within the garden of emergent hierarchies already shown in Figure 4, which is the “balloon"—where all the positive \(\Delta \mathrm{CP}\) in the entire emergent hierarchy is contained within an individual macroscale. Here we show two general methods for how construct such examples of pinpoint emergence. 

\begin{figure}[H]
    \centering
    \includegraphics[width=1\textwidth, height=0.8\textheight, keepaspectratio]{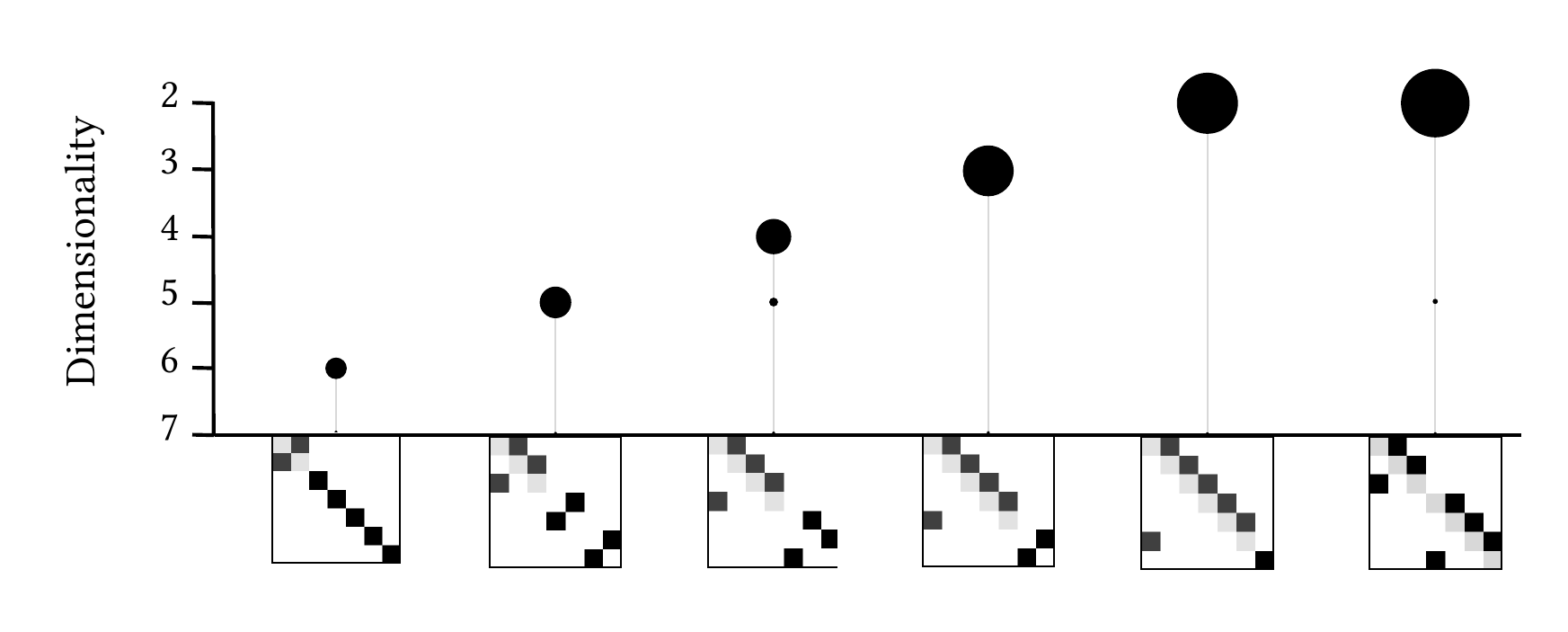}
    \caption{By introducing a set of cycles and/or permutations, the emergent scale can be engineered to be at any preferred dimensionality. Shown here are directed diffusion cycles with a probability of staying in a state or transitioning to the next point along the cycle of $0.2$ and $0.8$, respectively. \label{fig:balloons}}
\end{figure}

Specifically, we make use of the fact that a cycle of transitions has no hierarchical structure: there exists no intermediate scale between the individual nodes and the full cycle at which the dynamics simplify. Contracting the whole cycle, in contrast, yields a very efficient simplification, namely by representing the whole cycle as a single node with only a self-loop, drastically increasing $\text{CP}$. Figure \ref{fig:balloons} demonstrates examples of this phenomenon. To isolate the emergent scale at level $l$ (that is, the coarse-grained system has $l$ states), we introduce either a cycle of size $c\leq l$ and a set of $l-c$ permutations (possibly representing other levels that have already been simplified), or a total of $l$ disjoint cycles. The result is that, out of an astronomical number of scales of description (even the small systems in Figure \ref{fig:balloons} have a total of 877 possible scales), only one single macroscale contributes to the causal workings of the system.

\section{Discussion}

There are significant benefits to being able to engineer emergence. First, emergent systems may be more robust, in that they are more resistant to knockouts at the microscale \cite{klein2021evolution}, and also in that their macroscales are cryptic \cite{hoel2020emergence}. As we have also seen here, complex emergent hierarchies appear to have a close relationship with scale-freeness and criticality \cite{barabasi1999emergence, aldana2003natural}, even offering a new—quite literal—definition of scale-freeness as when no one scale dominates the causal workings of a system.

There has been increasing research on causal emergence (see Yuan et al. 2024 \cite{yuan2024emergence} for a review), based in the idea that macroscale causation can under some conditions be stronger (i.e., that macroscale causal relationships are more robust, less noisy, more necessary) and ultimately this stems from a kind of causal error-correction via multiple-realizability \cite{hoel2017map, hoel2024world} . Recently there has been the introduction of the theory of causal emergence 2.0, an update which is more theoretically and conceptually complete than its predecessor, and which negates previous mathematical criticisms of the theory (e.g., \cite{eberhardt2022causal}) that no longer apply to the new formalism; most importantly, as we demonstrate here, CE 2.0 distinguishes a taxonomy of different types of emergence, wherein a system's workings may be “top-heavy," “bottom-heavy," peak at a mesoscale, or be highly complex and so spread across the system's different levels (that is, at different degrees of dimension reduction). 

It is natural to, for various systems, desire different emergent outcomes. E.g., we may want a system to have maximal emergent complexity (and given the prevalence of classic scale-freeness in real data \cite{albert2002statistical}, we should expect real systems to also show high degrees of emergent complexity in the same manner). In other cases, it is natural to want certain systems to operate precisely and solely at particularly scales of description, such that only those scales, and those alone, drive the causal workings of the system. Important examples might be, for instance, engineering brain-computer interfaces to operate in synchrony with the relevant brain scales that correlate to human psychology or the neural correlates of consciousness \cite{crick1994astonishing, seth2022theories}. 

More generally, we posit it is a requisite for human freedom and autonomy that the scale of psychology (or the neural correlates of consciousness) is a contributing member of the emergent hierarchy of the brain and body, adding significantly to the system's total causal workings. Critically, we point out that the causal relevancy of human psychology must be preserved in any sort of AI-human merger, such as those being attempted via brain implants \cite{musk2019integrated}, and that the ability to engineer emergence may be necessary to prevent loss of human control, and that this area of research is currently acutely understudied. 

Several limitations of our paper are worth discussing, mostly in that it is focused on proof-of-principle results rather than large-scale empirical data. Even confined to systems of states we study, these are either small (so that we can brute force compare all the scales) and, even when analyzing the network growth rules, still not very large (only 40 states), so that we can easily run the branching greedy algorithm. However, in the future, we plan to apply this to larger networks, as currently our algorithm (detailed in SI Section 3) scales in subexponential time, and further speedups using parallelization are feasible. 

Throughout our analysis we use a $p(C)$ that is uniform for any TPM we consider, which reflects the assumption that each scale is treated independently during causal analysis (e.g., you don't count the atoms in a light switch to assess the impact of the intervention of $\text{\textit{do}(light switch = UP)}$ on a connected light bulb, you instead compare it to $\text{\textit{do}(light switch = DOWN)}$). A different choice would be to instead coarse-grain the microscopic $p(C)$; however, then the coarse-grained versions of $p(C)$ would merely correspond to an observational distribution based on interventions at a system's microscale (which violates how causation should go beyond observation) \cite{pearl2009causality}). To avoid this, we follow the equivalent of performing a randomized controlled trial (the heart of the scientific understanding of causation \cite{fisher1999advances}) at each scale to understand what its states do. While this is a defensible mathematical assumption for dealing with causation, it is important to note that, unlike its earlier version CE 1.0 \cite{eberhardt2022causal, aaronson_higher-level_2017, hoel2017map}, the mere existence of emergence in CE 2.0 does not depend on any specific assumptions regarding $p(C)$, e.g., such as using the uniform distribution. Indeed, gains in the causal primitives (and the many measures of causation based on them) can be identified across different assumptions of what $p(C)$ should be \cite{comolatti2025consilience}, and causal emergence can still occur even in cases where our assumption of scale-independence is violated, such as when the $p(C)$ used at a macroscale is purely observational and so is just a direct coarse-graining of the microscale $p(C)$  \cite{hoel2025causal}.

It's also important to note that not all macroscales herein are \textit{dynamically consistent} \cite{rubenstein2017causal}. Unlike in previous research on CE 1.0 and 2.0, we do not restrict our analysis to the set of scales which are dynamically consistent: i.e., such that the dynamics of the microscale are preserved (albeit coarse-grained) up at the macroscale. This is simply because we found that checking for dynamical consistency did not alter our results in a significant way (data not shown). 

Future research will explore opportunities for synthesis with existing theories of emergence or alternative or improved methods for how to partition or apportion causation across scales. For instance, in terms of causal emergence, there has been a number of efforts to provide alternative versions to CE 1.0, such as those based instead on information decomposition \cite{mediano2022greater} or dynamical reversibility \cite{zhang2025dynamical}. These methods will likely need to be expanded to better reflect the improvements of CE 2.0.  Additionally, there have recently been proposed novel methods for decomposing interventional causality  \cite{jansma2025decomposing}, as well as a new approach to Shapley values and Möbius inversions \cite{forré2025mobiustransformsshapleyvalues} that might serve as a future potential causal apportioning schema wherein the total $\text{CP}$ is preserved when summed across the sublattice.
 
Overall, our results show that CE 2.0 framework can be extended to account for multiple paths from the microscale to the macroscale, revealing the full emergent hierarchy of scales. This emergent hierarchy—the subset of all possible scales that actually causally contribute to the system's workings—has evocative and rich mathematical properties to explore, and gives significant information about the nature of emergence occurring in the system. Additionally, it opens the door for designing systems to possess desired emergent hierarchies, with important implications for science and engineering and the construction of causal models.

\section{Acknowledgments}

EH thanks Michael Levin for his encouragement, conversations, and feedback. AJ acknowledges support from the Dutch Institute for Emergent Phenomena (DIEP) cluster via the Foundations and Applications of Emergence (FAEME) programme.


\clearpage
\appendix
\renewcommand\thefigure{S\arabic{figure}}
\renewcommand\thetable{S\arabic{table}}
\renewcommand\thesection{S\arabic{section}}
\setcounter{figure}{0}
\setcounter{table}{0}
\setcounter{section}{0}

\section*{Supplementary Information}
\addcontentsline{toc}{section}{Supplementary Information}

\section{Ground-truth TPMs as causal models}
\label{SI:TPM_do_calc}
TPMs are often inferred from observational data with incomplete knowledge of the causal graph, and therefore generally do not encode causal effects. However, we here use them to \emph{encode} a causal model by assuming ground-truth knowledge, in that the system at time $t$ is fully characterized by a state $X_t$ (there are no hidden confounders), and the dynamics are Markovian, so that the (somewhat trivial) causal graph can be written as 

\begin{align}
    \ldots \to X_{t-1} \to X_t \to X_{t+1} \to X_{t+2} \to \ldots
\end{align}

Interventions of the type $do(X_t=x)$ correspond to putting the system in a certain state $x$. By the second rule of the do-calculus:
\begin{align}
    p(X_{t+1}|do(X_t=x)) = p(X_{t+1}|X_t=x) 
\end{align}
if $X_{t+1}$ is independent of $X_t$ in the graph with the outgoing arrows of $X_t$ removed. This is clearly true in this case:
\begin{align}
    \ldots \to X_{t-1} \to X_t  \quad \quad X_{t+1} \to X_{t+2} \to \ldots
\end{align}

Since a system's TPM is simply a way to summarize $p(X_{t+1}|X_t=x)$ for all $x$, we refer to such relationships as \emph{causal}. Since the states are always over the entirety of the system, we refer to any prior state as a \textit{cause}, and any of its immediately following states an an \textit{effect}.

For simplicity, we focus solely on the causal relationships between the states of a system (and their macroscales, in the form of sets of macrostates). Admittedly, such TPMs do not uniquely specify a given system's mechanisms or elements. Therefore, this work could be extended to be over causal models consisting of mechanisms and elements as well (e.g., as in \cite{hoel2013quantifying,marshall2018black}), rather than just causal models of states, although we feel for an initial treatment this adds unnecessary complexity (both conceptual and otherwise).

\section{Pseudocode of CE 2.0’s calculation across all possible scales of a system.}

\begin{algorithm}[H]
\DontPrintSemicolon
\SetKwInOut{Input}{Input}
\SetKwInOut{Output}{Output}
\caption{Calculating $\Delta$CP$(T)$ \label{alg:dCP}}

\Input{
  Row-stochastic matrix $T \in \mathbb{R}^{n \times n}$; \\
}
\Output{
  Hasse diagram $H$ over partitions; \\
  $\mathrm{CPdict}:\ \text{partition}\mapsto \mathbb{R}$; \\
  $\Delta\mathrm{CPdict}:\ \text{partition}\mapsto \mathbb{R}$; \\
  $\mathrm{Emergent}$: partitions with $\Delta\mathrm{CP}>\varepsilon$.

}

$n \gets$ number of states in $T$\;

\BlankLine
\textbf{1) Enumerate coarse-grainings}\;
$\mathcal{P} \gets \textsc{GenerateAllCoarseGrainings}(n)$\;


\BlankLine
\textbf{2) Compute CP on each partition}\;
$\mathrm{CPdict} \gets \emptyset$\;
\ForEach{$P \in \mathcal{P}$}{
  $\mathrm{CPdict}[P] \gets \textsc{CP}(\textsc{CoarseGrain}(T, P))$\;
}

\BlankLine
\textbf{3) Build refinement lattice (Hasse diagram)}\;
$H \gets \textsc{BuildRefinementGraph}(\mathcal{P})$\;

\BlankLine
\textbf{4) $\Delta$CP relative to ancestors}\;
$\Delta\mathrm{CPdict} \gets \emptyset$\;
\ForEach{$P \in \textsc{Nodes}(H)$}{
    $\mathcal{A} \gets \textsc{Ancestors}(H, P)$\;
  \uIf{$\mathcal{A}$ is empty}{
    $b \gets 0$
  }
  \Else{
    $b \gets \max\{\mathrm{CPdict}[Q] : Q \in \mathcal{A}\}$\;
  }
  $\Delta\mathrm{CPdict}[P] \gets \mathrm{CPdict}[P] - b$\;
}

\BlankLine
\textbf{5) Emergent set (positive $\Delta$CP)}\;
$\mathrm{Emergent} \gets \{\,P \in \mathcal{P} \mid \Delta\mathrm{CP}[P] > \varepsilon\,\}$\;

\BlankLine
\Return $(H,\ \mathrm{CP},\ \Delta\mathrm{CP},\ \mathrm{Emergent})$\;

\end{algorithm}

\section{A branching greedy algorithm for estimating the emergent hierarchy of a system.}

\begin{algorithm}[H]
\DontPrintSemicolon
\SetKwFunction{GreedyCompletion}{GreedyCompletion}
\SetKwProg{Fn}{Function}{:}{}
\caption{GreedyCompletion from a starting partition}
\Fn{\GreedyCompletion{$T$, $P_{\text{start}}$}}{
  $P \gets P_{\text{start}}$;\quad
  $\text{Path} \gets [P]$;\quad
  $\text{CPlist} \gets [\,\textsc{CP}(\textsc{CoarseGrain}(T,P))\,]$\;
  \While{$P$ has more than one block}{
    $k \gets$ number of blocks in $P$;\quad $(\text{best}, P^\star) \gets (-\infty, \bot)$\;
    \ForEach{unordered pair $(i,j)$ of block indices in $\{1,\dots,k\}$}{
      $Q \gets$ merge blocks $i$ and $j$ in $P$;\;
      $e \gets \textsc{CP}(\textsc{CoarseGrain}(T,Q))$;\;
      \If{$e > \text{best}$}{ $\text{best} \gets e$;\quad $P^\star \gets Q$ }
    }
    append $P^\star$ to \text{Path};\quad append \text{best} to \text{CPlist};\quad $P \gets P^\star$\;
  }
  \Return $(\text{Path}, \text{CPlist})$\;
}
\end{algorithm}

\begin{algorithm}[H]
\DontPrintSemicolon
\SetKwInOut{Input}{Input}
\SetKwInOut{Output}{Output}
\caption{$\Delta$CP$(T)$ from parallel greedy paths}
\Input{
  Row-stochastic matrix $T$;\quad
  \texttt{n\_paths} $\in \mathbb{N}$ (default $3$).
}
\Output{
  $H$: refinement DAG over sampled partitions;\\
  $\mathrm{CPdict}$: partition $\to$ CP;\\
  $\Delta\mathrm{CPdict}$ (via Alg.~1, Steps 4–5).
}

\textbf{1) Parallel greedy sampling of high-scoring branches}\;
$P_{\text{cur}} \gets \textsc{FinestGraining}(T)$\;
$\mathrm{CPdict} \gets \{\,P_{\text{cur}} \mapsto \textsc{CP}(T)\,\}$\\
\While{$P_{\text{cur}}$ has more than one block}{
  $k \gets$ number of blocks in $P_{\text{cur}}$;\\
  $\mathcal{C} \gets \emptyset$\\
  \ForEach{unordered pair $(i,j)$ of block indices in $\{1,\dots,k\}$}{
    $Q \gets$ merge blocks $i$ and $j$ in $P_{\text{cur}}$;\quad
    $e \gets \textsc{CP}(\textsc{CoarseGrain}(T,Q))$;\\
    append $(e,Q)$ to $\mathcal{C}$\;
  }
  $\text{Top} \gets$ the $\min(\texttt{n\_paths}, |\mathcal{C}|)$ pairs in $\mathcal{C}$ with largest $e$\;
  $\mathcal{S} \gets [\,Q : (e,Q) \in \text{Top}\,]$ \tcp*{selected starting merges}
  \ForEach{$Q \in \mathcal{S}$}{
    $(\text{Path}, \text{CPlist}) \gets \textsc{GreedyCompletion}(T, Q)$\;
    \For{$t=1$ \KwTo $|\text{Path}|$}{
      \If{\text{Path}[t] $\notin \mathrm{CPdict}$}{ $\mathrm{CP}[\text{Path}[t]] \gets \text{CPlist}[t]$ }
    }
  }
  $P_{\text{cur}} \gets$ first element of $\mathcal{S}$
}

\BlankLine
\textbf{2) Build refinement lattice over sampled nodes}\;
$H \gets \textsc{BuildRefinementGraph}(\text{keys}(\mathrm{CPdict}))$

\BlankLine
\textbf{3)} Apply \ref{alg:dCP}, Steps 4–5, with $(H,\mathrm{CPdict})$ to obtain $\Delta\mathrm{CPdict}$.\;
\end{algorithm}

\bibliographystyle{unsrt}  %
\bibliography{references}  %

\end{document}